\documentclass[envcountsame]{llncs}
\usepackage{amssymb}

\usepackage{times}

\title{Topological Complexity of  $\omega$-Powers : \\ Extended Abstract}
\author{Olivier Finkel\inst{1} \and Dominique Lecomte\inst{2}}
\institute{{\it Equipe Mod\`{e}les de Calcul et Complexit\'e}  
 \\ {\it Laboratoire de l'Informatique du Parall\'elisme}\footnote{UMR 5668 - CNRS - ENS Lyon - UCB Lyon - INRIA  \\ LIP Research Report RR 2008-27}
 \\  CNRS et Ecole Normale Sup\'erieure de Lyon
 \\ 46, All\'ee d'Italie 69364 Lyon Cedex 07, France. \\ \email{Olivier.Finkel@ens-lyon.fr} \and 
{\it Equipe d'Analyse 
Fonctionnelle}
\\ Universit\' e Paris 6
\\4, place Jussieu, 75 252 Paris Cedex 05, France
 \\ \email{dominique.lecomte@upmc.fr}}

\date{}
\begin{document}

\spnewtheorem{Rem}[theorem]{Remark}{\bfseries}{\itshape}
\spnewtheorem{Exa}[theorem]{Example}{\bfseries}{\itshape}

\spnewtheorem{prop}[theorem]{Proposition}{\bfseries}{\itshape}
\spnewtheorem{lem}[theorem]{Lemma}{\bfseries}{\itshape}
\spnewtheorem{cor}[theorem]{Corollary}{\bfseries}{\itshape}
\spnewtheorem{defi}[theorem]{Definition}{\bfseries}{\itshape}

\def\ufootnote#1{\let\savedthfn\thefootnote\let\thefootnote\relax
\footnote{#1}\let\thefootnote\savedthfn\addtocounter{footnote}{-1}}

\newcommand{\bormxi}{{\bf\Pi}^{0}_{\xi}}
\newcommand{\bormlxi}{{\bf\Pi}^{0}_{<\xi}}
\newcommand{\bormz}{{\bf\Pi}^{0}_{0}}
\newcommand{\bormone}{{\bf\Pi}^{0}_{1}}
\newcommand{\ca}{{\bf\Pi}^{1}_{1}}
\newcommand{\bormtwo}{{\bf\Pi}^{0}_{2}}
\newcommand{\bormthree}{{\bf\Pi}^{0}_{3}}
\newcommand{\bormom}{{\bf\Pi}^{0}_{\omega}}
\newcommand{\borom}{{\bf\Delta}^{0}_{\omega}}
\newcommand{\borml}{{\bf\Pi}^{0}_{\lambda}}
\newcommand{\bormlpn}{{\bf\Pi}^{0}_{\lambda +n}}
\newcommand{\bormpm}{{\bf\Pi}^{0}_{1+m}}
\newcommand{\borapm}{{\bf\Sigma}^{0}_{1+m}}
\newcommand{\bormep}{{\bf\Pi}^{0}_{\eta +1}}
\newcommand{\borapxi}{{\bf\Sigma}^{0}_{\xi}}
\newcommand{\borai}{{\bf\Sigma}^{0}_{ 2.\xi +1 }}
\newcommand{\bormpxi}{{\bf\Pi}^{0}_{\xi}}
\newcommand{\bormpeta}{{\bf\Pi}^{0}_{1+\eta}}
\newcommand{\borapxipo}{{\bf\Sigma}^{0}_{\xi +1}}
\newcommand{\bormpxipo}{{\bf\Pi}^{0}_{\xi +1}}
\newcommand{\borpxi}{{\bf\Delta}^{0}_{\xi}}
\newcommand{\borel}{{\bf\Delta}^{1}_{1}}
\newcommand{\Borel}{{\it\Delta}^{1}_{1}}
\newcommand{\borone}{{\bf\Delta}^{0}_{1}}
\newcommand{\bortwo}{{\bf\Delta}^{0}_{2}}
\newcommand{\borthree}{{\bf\Delta}^{0}_{3}}
\newcommand{\boraone}{{\bf\Sigma}^{0}_{1}}
\newcommand{\boratwo}{{\bf\Sigma}^{0}_{2}}
\newcommand{\borathree}{{\bf\Sigma}^{0}_{3}}
\newcommand{\boraom}{{\bf\Sigma}^{0}_{\omega}}
\newcommand{\boraxi}{{\bf\Sigma}^{0}_{\xi}}
\newcommand{\ana}{{\bf\Sigma}^{1}_{1}}
\newcommand{\pca}{{\bf\Sigma}^{1}_{2}}
\newcommand{\Ana}{{\it\Sigma}^{1}_{1}}
\newcommand{\Boraone}{{\it\Sigma}^{0}_{1}}
\newcommand{\Borone}{{\it\Delta}^{0}_{1}}
\newcommand{\Bormone}{{\it\Pi}^{0}_{1}}
\newcommand{\Bormtwo}{{\it\Pi}^{0}_{2}}
\newcommand{\Ca}{{\it\Pi}^{1}_{1}}
\newcommand{\bormn}{{\bf\Pi}^{0}_{n}}
\newcommand{\bormm}{{\bf\Pi}^{0}_{m}}
\newcommand{\boralp}{{\bf\Sigma}^{0}_{\lambda +1}}
\newcommand{\borat}{{\bf\Sigma}^{0}_{|\theta |}}
\newcommand{\bormat}{{\bf\Pi}^{0}_{|\theta |}}
\newcommand{\Borapxi}{{\it\Sigma}^{0}_{\xi}}
\newcommand{\Bormpxipo}{{\it\Pi}^{0}_{1+\xi +1}}
\newcommand{\Borapn}{{\it\Sigma}^{0}_{1+n}}
\newcommand{\borapn}{{\bf\Sigma}^{0}_{1+n}}
\newcommand{\boraxipm}{{\bf\Sigma}^{0}_{\xi^\pm}}
\newcommand{\Boratwo}{{\it\Sigma}^{0}_{2}}
\newcommand{\Borathree}{{\it\Sigma}^{0}_{3}}
\newcommand{\Borapnpo}{{\it\Sigma}^{0}_{1+n+1}}
\newcommand{\Bormpxi}{{\it\Pi}^{0}_{\xi}}
\newcommand{\Borpxi}{{\it\Delta}^{0}_{\xi}}
\newcommand{\boratpxi}{{\bf\Sigma}^{0}_{2+\xi}}
\newcommand{\Boratpxi}{{\it\Sigma}^{0}_{2+\xi}}
\newcommand{\bormltpxi}{{\bf\Pi}^{0}_{<2+\xi}}
\newcommand{\Bormltpxi}{{\it\Pi}^{0}_{<2+\xi}}
\newcommand{\borapeap}{{\bf\Sigma}^{0}_{1+\eta_{\alpha ,p}}}
\newcommand{\borapeapn}{{\bf\Sigma}^{0}_{1+\eta_{\alpha ,p,n}}}
\newcommand{\Borapeap}{{\it\Sigma}^{0}_{1+\eta_{\alpha ,p}}}
\newcommand{\Bormpn}{{\it\Pi}^{0}_{1+n}}
\newcommand{\Borpn}{{\it\Delta}^{0}_{1+n}}
\newcommand{\borapximo}{{\bf\Sigma}^{0}_{1+(\xi -1)}}
\newcommand{\borpeta}{{\bf\Delta}^{0}_{1+\eta}}

\newcommand{\hs}{\hspace{12mm}

\noi}
\newcommand{\noi}{\noindent}
\newcommand{\ol}{ $\omega$-language}
\newcommand{\om}{\omega}
\newcommand{\Si}{\Sigma}
\newcommand{\Sis}{\Sigma^\star}
\newcommand{\Sio}{\Sigma^\omega}
\newcommand{\nl}{\newline}
\newcommand{\lra}{\leftrightarrow}
\newcommand{\fa}{\forall}
\newcommand{\ra}{\rightarrow}
\newcommand{\orl}{ $\omega$-regular language}

\maketitle


\section{Introduction}

\noi
The operation $V \ra V^\om$  is a fundamental operation over finitary languages 
leading to \ol s. It produces $\omega$-powers, i.e.   \ol s in the form $V^\om$, where $V$ is a finitary language. 
This operation  appears  in the characterization of the class 
$REG_\om$  of  \orl s (respectively,  of the class $CF_\om$ of context free \ol s) 
 as the $\om$-Kleene closure 
of the family $REG$ of regular finitary languages (respectively,   of the 
family $CF$ of context free finitary languages) \cite{sta}.  
\nl Since the set $\Sio$ of infinite words over a finite alphabet $\Si$ can be   equipped 
with the usual Cantor topology, the question of  the topological  complexity of  $\om$-powers of 
finitary  languages naturally arises and has  been posed by 
Niwinski \cite{Niwinski90},  Simonnet \cite{Simonnet92},  and  Staiger \cite{sta}. 
A first task is to study the position of $\om$-powers 
with regard to the Borel hierarchy (and beyond to the projective hierarchy) \cite{sta,pp}. 

\hs It is easy to see that the  $\om$-power of a finitary language is always an analytic set because 
it is either the continuous image of a compact set $\{0,1, \ldots ,n\}^\om$ for $n\geq 0$
or of the Baire space $\om^\om$. 

\hs  It has been recently  proved,   that 
for each integer $n\geq 1$, there exist some $\om$-powers of  context free languages 
which are ${\bf \Pi}_n^0$-complete Borel sets,  \cite{Fin01a},  and   that there exists a 
context free language $L$ such that $L^\om$ is analytic but not Borel, \cite{Fin03a}. Notice that amazingly 
the language $L$ is  very simple to describe and it is accepted by a simple $1$-counter automaton. 

\hs  The first author proved in \cite{Fin04} that  there exists a finitary language $V$ 
such that $V^\om$ is a Borel set of infinite rank. It was also proved in   \cite{DF06} that 
 there is a context free language $W$ such that $W^\om$ is Borel above 
${\bf \Delta_\omega^0}$. 
\nl 
We  proved  in \cite{Fin-Lec}  the following very surprising result which shows that $\omega$-powers exhibit a great topological complexity: 
 for each non-null countable ordinal  $\xi$, there  exist 
some $\borapxi$-complete $\omega$-powers, and some  $\bormpxi$-complete  $\omega$-powers. 

\hs We consider also the Wadge hierarchy which is a great refinement of the Borel hierarchy. We get  
 many more Wadge degrees of $\omega$-powers, showing that for each ordinal $\xi \geq 3$, there are uncountably many Wadge degrees of 
$\omega$-powers of Borel rank $\xi +1$. 

\hs We show  also, using some tools of effective descriptive set theory,  that the main result of \cite{Fin-Lec} has some effective counterparts.  
 
\hs All the proofs of the results presented here may be found  in the conference paper \cite{Fin-Lec} or in the preprint \cite{Fin-Lec2} which contains 
also some additional results. 

\section{Topology}

\noi We first give some notations for finite or infinite words, assuming  
the reader to be familiar with the theory of formal languages and of 
$\om$-languages, see \cite{tho,sta,pp}.  
\noi Let $\Si$ be a finite or countable alphabet whose elements are called letters.
A non-empty finite word over $\Si$ is a finite sequence of letters:
 $x=a_0.a_1.a_2\ldots a_n$ where $\fa i\in [0; n]$ $a_i \in\Si$.
 We shall denote $x(i)=a_i$ the $(i+1)^{th}$ letter of $x$. The length of $x$ is $|x|=n+1$.
The empty word  has 0 letters. Its length is 0.
 The set of finite words over $\Si$ is denoted $\Si^{<\om}$.
 A (finitary) language $L$ over $\Si$ is a subset of $\Si^{<\om}$.
 The usual concatenation product of $u$ and $v$ will be denoted by $u^\frown v$ or just  $uv$.

\hs The first infinite ordinal is $\om$.
An $\om$-word over $\Si$ is an $\om$ -sequence $a_0a_1\ldots a_n \ldots$, where 
for all integers  $i\geq 0$~~ $a_i \in\Sigma$.  
 When $\sigma$ is an $\om$-word over $\Si$, we write
 $\sigma =\sigma(0)\sigma(1)\ldots  \sigma(n) \ldots $.
The set of $\om$-words over  the alphabet $\Si$ is denoted by $\Si^\om$.
 An  $\om$-language over an alphabet $\Sigma$ is a subset of  $\Si^\om$. 
 The concatenation product is also extended to the product of a 
finite word $u$ and an $\om$-word $v$: 
the infinite word $u.v$ or $u^\frown v$ is then the $\om$-word such that:
 $(u v)(k)=u(k)$  if $k < |u|$ , and  $(u.v)(k)=v(k-|u|)$  if $k\geq |u|$.
\nl 
 The prefix relation is denoted $\prec$: the finite word $u$ is a prefix of the finite 
word $v$ (respectively,  the infinite word $v$), denoted $u \prec v$,  
 if and only if there exists a finite word $w$ 
(respectively,  an infinite word $w$), such that $v=u^\frown  w$.

\hs For a finitary language $V \subseteq \Si^{<\om}$, the $\om$-power of $V$ is the $\om$-language 
 $$V^\om = \{ u_1\ldots  u_n\ldots  \in \Si^\om \mid  \fa i\geq 1 ~~ u_i\in V \}$$

\noi We recall now some notions of topology, assuming  the reader to be familiar with basic notions  which
may be found in  \cite{ku,mos,kec,lt,sta,pp}.
\nl There is a natural metric on the set $\Sio$ of  infinite words 
over a countable alphabet 
$\Si$ which is called the prefix metric and defined as follows. For $u, v \in \Sio$ and 
$u\neq v$ let $d(u, v)=2^{-l_{pref(u,v)}}$ where $l_{pref(u,v)}$ is the first integer $n$
such that the $(n+1)^{th}$ letter of $u$ is different from the $(n+1)^{th}$ letter of $v$. 
The topology induced on  $\Sigma^\omega$ by this metric is just
 the product topology of the discrete topology on $\Sigma$. 
 For $s \in \Si^{<\om}$, the set 
$N_{s}\! :=\!\{\alpha\!\in\!\Sigma^\omega\mid s\!\prec\!\alpha\}$ 
is a basic clopen (i.e., closed and open) set of $\Sigma^\omega$. 
More generally open sets of $\Sio$ are in the form $W^\frown \Si^\om$, where $W\subseteq \Si^{<\om}$.

\hs 
 When $\Si$ is a finite alphabet, 
the prefix  metric induces on $\Sio$ the usual  Cantor topology and $\Sio$ is compact. 
\nl The Baire space $\omega^\omega$ is equipped with the product topology of the 
discrete topology on $\omega$. It is homeomorphic to $P_\infty\! :=\!\{\alpha\!\in\! 2^\omega\mid
\forall i\!\in\!\omega\ \exists j\!\geq\! i\ \ \alpha (j)\! =\! 1\}\!\subseteq\! 2^\omega$, via the map 
defined on $\omega^\omega$ by $H(\beta )\! :=\! 0^{\beta (0)}10^{\beta (1)}1 \ldots$

\hs We define now the  {\bf Borel Hierarchy} on a  topological space $X$:

\begin{defi}
The classes ${\bf \Si}_\xi^0(X)$ and ${\bf \Pi}_\xi^0(X) $ of the Borel Hierarchy
 on the topological space  $X$ are defined as follows:
\nl ${\bf \Si}^0_1(X) $ is the class of open subsets of $X$.
\nl ${\bf \Pi}^0_1(X) $ is the class of closed subsets of $X$.
\nl And for any countable ordinal $\xi \geq 2$:
\nl ${\bf \Si}^0_\xi (X)$ is the class of countable unions of subsets of $X$ in 
$\cup_{\gamma <\xi}{\bf \Pi}^0_\gamma $.
 \nl ${\bf \Pi}^0_\xi (X)$ is the class of countable intersections of subsets of $X$ in 
$\cup_{\gamma <\xi}{\bf \Si}^0_\gamma $.
\nl As usual  the ambiguous class  $\borpxi$ is the class $\borapxi\cap\bormpxi$.
\end{defi}

\noi  Suppose now that $X\!\subseteq\! Y$; then  
$\borapxi (X)\! =\!\{ A\cap X\mid A\!\in\!\borapxi (Y)\}$, and similarly for $\bormpxi$, see   \cite[Section 22.A]{kec}.
Notice that we have defined the Borel classes  ${\bf \Si}^0_\xi(X)$ and ${\bf \Pi}^0_\xi (X)$ mentioning the space $X$. 
However when the context is clear we will sometimes omit $X$ and denote ${\bf \Si}^0_\xi(X)$ by ${\bf \Si}^0_\xi$ and 
similarly for the dual class.

\hs The class of 
{\bf Borel\ sets} is $\borel\! :=\!\bigcup_{\xi <\omega_1}\ \borapxi\! =\!
\bigcup_{\xi <\omega_1}\ \bormpxi$, where $\om_1$ is the first uncountable ordinal. 

\hs For 
a countable ordinal $\alpha$,  a subset of $\Si^\om$ is a Borel set of {\it rank} $\alpha$ iff 
it is in ${\bf \Si}^0_{\alpha}\cup {\bf \Pi}^0_{\alpha}$ but not in 
$\bigcup_{\gamma <\alpha}({\bf \Si}^0_\gamma \cup {\bf \Pi}^0_\gamma)$.

\hs  We now define completeness with regard to reduction by continuous functions. 
For a countable ordinal  $\alpha\geq 1$, a set $F\subseteq \Si^\om$ is said to be 
a ${\bf \Si}^0_\alpha$  
(respectively,  ${\bf \Pi}^0_\alpha$)-{\it complete set} 
iff for any set $E\subseteq Y^\om$  (with $Y$ a finite alphabet): 
 $E\in {\bf \Si}^0_\alpha$ (respectively,  $E\in {\bf \Pi}^0_\alpha$) 
iff there exists a continuous function $f: Y^\om \ra \Si^\om$ such that $E = f^{-1}(F)$. 
  ${\bf \Si}^0_n$
 (respectively, ${\bf \Pi}^0_n$)-complete sets, with $n$ an integer $\geq 1$, 
 are thoroughly characterized in \cite{stac}.  
\nl Recall that 
a set $X \subseteq \Sio$ is a ${\bf \Si}^0_\alpha$
 (respectively ${\bf \Pi}^0_\alpha$)-complete subset of $\Sio$ iff it is in ${\bf \Si}^0_\alpha$ 
but not in ${\bf \Pi^0_\alpha}$  (respectively in  ${\bf \Pi}^0_\alpha$ but not in  ${\bf \Si}^0_\alpha$), \cite{kec}. 

\hs 
For example, the singletons of $2^\omega$ are $\bormone$-complete subsets of $2^\omega$. 
The set $P_\infty$ is a well known example of a $\bormtwo$-complete subset of $2^\omega$.

\hs  If ${\bf\Gamma}$ is a class of sets, then $\check {\bf\Gamma}\! :=\!\{\neg A\mid A\!\in\! {\bf\Gamma}\}$
 is the class 
of complements of sets in ${\bf\Gamma}$. In particular, for every 
 non-null countable ordinal $\alpha$, $\check {{\bf \Si}^0_\alpha}\!= {\bf \Pi}^0_\alpha$ and 
$\check {{\bf \Pi}^0_\alpha}\!= {\bf \Si}^0_\alpha$.  

\hs We now introduce the Wadge hierarchy, which is a great refinement of the Borel hierarchy defined 
via reductions by continuous functions, \cite{Wadge83,Duparc01}. 

\begin{defi}[Wadge \cite{Wadge83}] Let $X$, $Y$ be two finite alphabets. 
For $L\subseteq X^\om$ and $L'\subseteq Y^\om$, $L$ is said to be Wadge reducible to $L'$
($L\leq _W L')$ iff there exists a continuous function $f: X^\om \ra Y^\om$, such that
$L=f^{-1}(L')$.
\nl $L$ and $L'$ are Wadge equivalent iff $L\leq _W L'$ and $L'\leq _W L$. 
This will be denoted by $L\equiv_W L'$. And we shall say that 
$L<_W L'$ iff $L\leq _W L'$ but not $L'\leq _W L$.
\nl  A set $L\subseteq X^\om$ is said to be self dual iff  $L\equiv_W L^-$, and otherwise 
it is said to be non self dual.
\end{defi}

\noi
 The relation $\leq _W $  is reflexive and transitive,
 and $\equiv_W $ is an equivalence relation.
\nl The {\it equivalence classes} of $\equiv_W $ are called {\it Wadge degrees}. 
\nl The Wadge hierarchy $WH$ is the class of Borel subsets of a set  $X^\om$, where  $X$ is a finite set,
 equipped with $\leq _W $ and with $\equiv_W $.
\nl  For $L\subseteq X^\om$ and $L'\subseteq Y^\om$, if   
$L\leq _W L'$ and $L=f^{-1}(L')$  where $f$ is a continuous 
function from $ X^\om$  into $Y^\om$, then $f$ is called a continuous reduction of $L$ to 
$L'$. Intuitively it means that $L$ is less complicated than $L'$ because 
to check whether $x\in L$ it suffices to check whether $f(x)\in L'$ where $f$ 
is a continuous function. Hence the Wadge degree of an \ol~
is a measure 
of its topological complexity. 
\nl
Notice  that in the above definition, we consider that a subset $L\subseteq  X^\om$ is given
together with the alphabet $X$.

\noi We can now define the {\it Wadge class} of a set $L$:

\begin{defi}
Let $L$ be a subset of $X^\om$. The Wadge class of $L$ is :
$$[L]= \{ L' \mid  L'\subseteq Y^\om \mbox{ for a finite alphabet }Y   \mbox{  and  } L'\leq _W L \}.$$ 
\end{defi}

\noi Recall that each {\it Borel class} ${\bf \Si^0_\alpha}$ and ${\bf \Pi^0_\alpha}$ 
is a {\it Wadge class}. 
A set $L\subseteq X^\om$ is a ${\bf \Si^0_\alpha}$
 (respectively ${\bf \Pi^0_\alpha}$)-{\it complete set} iff for any set 
$L'\subseteq Y^\om$, $L'$ is in 
${\bf \Si^0_\alpha}$ (respectively ${\bf \Pi^0_\alpha}$) iff $L'\leq _W L $ .
 
\begin{theorem} [Wadge]\label{wh}
Up to the complement and $\equiv _W$, the class of Borel subsets of $X^\om$,
 for  a finite alphabet $X$, is a well ordered hierarchy.
 There is an ordinal $|WH|$, called the length of the hierarchy, and a map
$d_W^0$ from $WH$ onto $|WH|-\{0\}$, such that for all $L, L' \subseteq X^\om$:
\nl $d_W^0 L < d_W^0 L' \lra L<_W L' $  and 
\nl $d_W^0 L = d_W^0 L' \lra [ L\equiv_W L' $ or $L\equiv_W L'^-]$.
\end{theorem}

\noi 
 The Wadge hierarchy of Borel sets of {\bf finite rank }
has  length $^1\varepsilon_0$ where $^1\varepsilon_0$
 is the limit of the ordinals $\alpha_n$ defined by $\alpha_1=\om_1$ and 
$\alpha_{n+1}=\om_1^{\alpha_n}$ for $n$ a non negative integer, $\om_1$
 being the first non countable ordinal. Then $^1\varepsilon_0$ is the first fixed 
point of the ordinal exponentiation of base $\om_1$. The length of the Wadge hierarchy 
of Borel sets in ${\bf \Delta^0_\om}= {\bf \Si^0_\om}\cap {\bf \Pi^0_\om}$ 
  is the $\om_1^{th}$ fixed point 
of the ordinal exponentiation of base $\om_1$, which is a much larger ordinal. The length 
of the whole Wadge hierarchy of Borel sets is a huge ordinal, with regard 
to the $\om_1^{th}$ fixed point 
of the ordinal exponentiation of base $\om_1$. It is described in \cite{Wadge83,Duparc01} 
by the use of the Veblen functions.

\hs There are some subsets of the topological 
space $\Sio$   which are not Borel sets. In particular, there 
exists another hierarchy beyond the Borel hierarchy,  called the 
projective hierarchy. 
 The first class of the projective hierarchy is the class 
 ${\bf \Si}^1_1$ of {\bf analytic} sets. 
A set $A \subseteq \Sio$ is analytic 
iff there exists a Borel set $B \subseteq (\Si \times Y)^\om$, with $Y$  a finite alphabet,  
such that $ x \in A \lra \exists y \in Y^\om $ such that $(x, y) \in B$, 
where $(x, y)\in (\Si \times Y)^\om$ is defined by:  
$(x, y)(i)=(x(i),y(i))$ for all integers $i\geq 0$.  
\nl
A subset of $\Sigma^\omega$ is  analytic if it is empty, 
or the image of the Baire space by a continuous map. 
The class of analytic sets contains the class of Borel sets in any of the spaces $\Sigma^\omega$. 
Notice that ${\bf \Delta}_1^1 = {\bf \Si}^1_1 \cap {\bf \Pi}^1_1$, where ${\bf \Pi}^1_1$ is the class of co-analytic sets, i.e. of complements of 
analytic sets. 
 
\hs The $\om$-power of a finitary language $V$ is always an analytic set because if $V$ is finite and has $n$ elements then 
$V^\om$  is the continuous image of a compact set $\{0,1, \ldots ,n-1\}^\om$ and if $V$ is infinite then 
there is a bijection  between $V$ and $\om$ and $V^\om$  is the continuous image 
 of the Baire space $\om^\om$, 
\cite{Simonnet92}.

\section{Topological complexity of  $\omega$-powers }

\noi  We now state our first main result, showing that $\om$-powers exhibit a very surprising  topological complexity.

\begin{theorem}[\cite{Fin-Lec}]\label{main-result}
 Let $\xi$ be a non-null countable ordinal.\smallskip  

\noindent (a) There is $A\!\subseteq\! 2^{<\omega}$ such that $A^\om$ is $\borapxi$-complete.\smallskip

\noindent (b) There is $A\!\subseteq\! 2^{<\omega}$ such that $A^\om$ is 
$\bormpxi$-complete.

\end{theorem}
 
\noi To prove Theorem \ref{main-result},  we  use in \cite{Fin-Lec} a level by level version of a theorem  of Lusin and Souslin 
stating that every Borel set $B \subseteq 2^\om$ is the image of a closed subset of the Baire space 
$\om^\om$ by a continuous bijection, see \cite[p.83]{kec}. 
It is the following theorem, proved by Kuratowski in \cite[Corollary 33.II.1]{ku}:
 
\begin {theorem}\label{kur}
 Let $\xi$ be a non-null countable ordinal, and $B\!\in\!\bormpxipo (2^\omega )$. 
Then there is $C\!\in\!\bormone (\omega^\omega )$ and a continuous bijection 
$f\! :\! C\!\rightarrow\! B$ such that $f^{-1}$ is $\borapxi$-measurable (i.e., $f[U]$ is $\borapxi (B)$ for each 
open subset $U$ of $C$).
\end{theorem} 

\noi The existence of the continuous bijection 
$f\! :\! C\!\rightarrow\! B$ 
given by this theorem (without the fact
that $f^{-1}$ is $\borapxi$-measurable) has been used by Arnold in \cite{Arnold83} to prove that every Borel subset of 
$\Sio$, for a finite alphabet $\Si$,  is accepted by a non-ambiguous finitely branching  transition system with B\"uchi acceptance   
condition. Notice that the sets of states of these transition systems are countable.
\nl Our first idea was to code the behaviour of such a transition system. In fact this can be done on a part of $\om$-words of a special 
compact set $K_{0,0}$. However  we  have also to consider more general sets     $K_{N,j}$ and then we  need the hypothesis of 
the $\borapxi$-measurability of the function $f$. 
 The complete proof can be found in \cite{Fin-Lec,Fin-Lec2}.

\hs  Notice that for   the class $\boratwo$, we need another proof, which uses a new operation which 
 is very close to the erasing operation defined by Duparc in his study of the Wadge hierarchy, \cite{Duparc01}. We get the following result. 

\begin{theorem}\label{th2}
 There is a context-free  language $A\!\subseteq\! 2^{<\omega}$ such that 
$A^\om\!\in\!\boratwo\!\setminus\!\bormtwo$.
\end{theorem}

\noi  Notice that it is easy to see that the set $2^\omega \setminus P_\infty$, which is  the classical example of $\boratwo$-complete set,   is not an $\om$-power. 
The question is still open to know whether there exists a {\it regular} language $L$ such that $L^\om$ is $\boratwo$-complete. 

\hs Recall that, for each  non-null countable ordinal $\xi$, the class of  $\borapxi$-complete (respectively,  $\bormpxi$-complete) subsets of 
$2^\om$ forms a single  {\it non self-dual} Wadge degree. 
 Thus Theorem \ref{main-result} provides also some Wadge degrees of $\om$-powers. More generally, 
it is natural to ask for the Wadge hierarchy of $\om$-powers. 
In the long version \cite{Fin-Lec2} of  the conference paper \cite{Fin-Lec} we get many more Wadge degrees of $\om$-powers. 

\hs In order to state these new results, we now recall the notion of difference hierarchy.  
(Recall that a countable ordinal $\gamma$ is said to be even iff it can be written in the form 
$\gamma=\alpha + n$, where $\alpha$ is a limit ordinal and $n$ is an even positive integer; otherwise the ordinal $\gamma$ is said to be odd; notice that all 
limit ordinals are even ordinals.)

\hs  If $\eta\! <\!\omega_1$ and $(A_\theta )_{\theta <\eta}$ is an increasing sequence of subsets of some space $X$, then we set 
$$D_\eta [(A_\theta )_{\theta <\eta}]\! :=\!\{ x\!\in\! X\mid\exists\theta\! <\!\eta~\ \ 
x\!\in\! A_\theta\!\setminus\!\bigcup_{\theta'<\theta}\ A_{\theta'}\ \ \hbox{\rm and\ the\ parity\ of}\ \ \theta\ \ \hbox{\rm is opposite\ to\ that\ of}\ \ \eta\}.$$
If moreover ~ $1\!\leq\!\xi\! <\!\omega_1$, ~~then we set : 

$$D_\eta (\boraxi )\! :=\!\{ D_\eta 
[(A_\theta )_{\theta <\eta}]\mid \mbox{ for each }  \theta <\eta ~~ A_\theta  \mbox{ is in the class } \boraxi\}.$$ 

\hs Recall that for each non null countable ordinal $\xi$, the sequence $( D_\eta (\boraxi ) )_{\eta < \om_1}$ is strictly increasing for the inclusion relation and that 
for each $\eta < \om_1$ it holds that $D_\eta (\boraxi ) \subseteq {\bf \Delta}^0_{\xi +1}$.  
Moreover for each $\eta < \om_1$ the class $D_\eta (\boraxi )$ is a Wadge class and  the class of  $D_\eta (\boraxi )$-complete subsets of 
$2^\om$ forms a single {\it non self-dual} Wadge degree. 

\begin{theorem}
\noi
\begin{enumerate}
\item  Let $1\!\leq\!\xi\! <\!\omega_1$. Then there is 
$A\!\subseteq\! 2^{<\omega}$ such that $A^\om$ is $\check D_2 (\boraxi )$-complete.

\item Let $3\!\leq\!\xi\! <\!\omega_1$ and $1 \leq \theta < \omega_1$. Then there is $A\!\subseteq\! 2^{<\omega}$ such that $A^\om$ is 
$\check D_{\om^\theta}(\boraxi )$-complete.
\end{enumerate}
\end{theorem}

\noi Notice that for each ordinal $\xi$ such that  $3\!\leq\!\xi\! <\!\omega_1$  we get uncountably many Wadge degrees of $\om$-powers of 
the same Borel rank $\xi +1$. 
This confirms the great complexity of these $\om$-languages. 
\nl However  the problem is still open to determine completely the Wadge hierarchy of $\om$-powers. 

\hs We now come to the effectiveness questions. 
It is natural to wonder 
whether the $\om$-powers  obtained above  are effective. For instance could they be obtained as $\om$-powers  of recursive languages ? 

\hs In the paper  \cite{Fin-Lec2} we prove effective versions of the results presented above. Using 
tools  of effective descriptive set theory, such Kleene recursion Theorem and the notion of Borel codes, 
 we first prove an effective version of Kuratowski's Theorem \ref{kur}. Then we use it 
to prove the following  effective version 
of Theorem \ref{main-result}, where $\Borapxi$ and $\Bormpxi$ denote classes of the hyperarithmetical hierarchy and $\omega_1^{CK}$ is the first 
non-recursive ordinal,  usually called the Church-kleene ordinal.

\begin{theorem}\label{recursive} 
 Let $\xi\!$ be a non-null ordinal smaller than $\omega_1^{CK}$.\smallskip 

\noindent (a) There is a recursive language $A\!\subseteq\! 2^{<\omega}$ such that 
$A^\om\!\in\!\Borapxi\!\setminus\!\bormpxi$.\smallskip

\noindent (b) There is a recursive language $A\!\subseteq\! 2^{<\omega}$ such that $A^\om\!\in\!\Bormpxi\!\setminus\!\borapxi$.\smallskip

\end{theorem}

\begin{Rem}
If $A\!\subseteq\! 2^{<\omega}$ is a recursive language, then the $\om$-power $A^\om$  is an effective analytic set, i.e. a (lightface) $\Si^1_1$-set. 
And the supremum of the set of Borel ranks of Borel effective analytic sets is the ordinal 
 $\gamma_2^1$. 
This ordinal is defined by   Kechris, Marker, and   Sami in 
\cite{kms} and it is proved to be strictly greater than the ordinal 
$\delta_2^1$ which is the first non $\Delta_2^1$ ordinal. Thus the ordinal  $\gamma_2^1$ is also strictly greater than the first non-recursive 
ordinal  $ \om_1^{\mathrm{CK}}$.
Thus Theorem \ref{recursive} does not give the complete answer about the Borel hierarchy of $\om$-powers of recursive languages. 
Indeed there could exist some $\om$-powers of recursive languages of Borel ranks greater than $\omega_1^{CK}$, but of course smaller than the 
ordinal  $\gamma_2^1$. 

\end{Rem}

\section{Concluding remarks}

\noi  The question  naturally arises 
to know what are all the possible infinite Borel ranks of 
$\om$-powers of  finitary  
languages belonging to some natural class like the 
class of context free languages (respectively,   languages 
accepted by stack automata,  recursive languages, 
 recursively enumerable languages,  \ldots ). 
\nl We know from \cite{mscs06} that there are $\om$-languages accepted by B\"uchi $1$-counter automata of every Borel rank (and even of every 
Wadge degree) of an effective analytic set.  Every $\om$-language accepted by a B\"uchi $1$-counter automaton can be written as a finite union 
$L = \bigcup_{1\leq i\leq n} U_i^\frown V_i^\om$, where for each integer $i$, $U_i$ and $V_i$ are finitary languages accepted by $1$-counter automata. 
And the supremum of the set of Borel ranks of effective analytic sets is the ordinal 
 $\gamma_2^1$.
From these results it seems  plausible that there exist some $\om$-powers of  languages accepted by  $1$-counter automata which have 
Borel ranks up to the ordinal  $\gamma_2^1$, although these languages are located at the very low level in the complexity hierarchy of finitary languages. 

\hs Another interesting question would be  to determine completely the 
Wadge hierarchy of $\om$-powers. A simpler open question is to determine the Wadge hierarchy of $\om$-powers of regular languages. 
The second author has given in \cite{Lecomte05} a few  Wadge degrees of  $\om$-powers of regular languages. 
Notice however that even  the question to determine the Wadge degrees of  $\om$-powers of regular languages in the class ${\bf \Delta}_2^0$ is still open.


\begin{thebibliography}{99}

\bibitem[Arn83]{Arnold83}
A. Arnold, Topological Characterizations of Infinite Behaviours 
of Transition Systems,  Automata, Languages and Programming, J. Diaz Ed., 
Lecture Notes in Computer Science, Volume154, Springer, 1983, p. 28-38. 


\bibitem[ABB96]{ABB96} J-M. Autebert, J. Berstel and L. Boasson, Context Free Languages
 and Pushdown Automata, in Handbook of Formal Languages, Vol 1, Springer Verlag 1996.

\bibitem[Dup01]{Duparc01} J. Duparc, Wadge Hierarchy and Veblen Hierarchy: Part 
1: Borel Sets of Finite Rank,  Journal of Symbolic Logic, Vol. 66, no. 1, 2001, p. 56-86.

\bibitem[DF07]{DF06} J. Duparc and O. Finkel, 
An  $\omega$-Power of a  Context-Free Language  Which Is Borel Above
$\Delta^0_\omega$, in  the Proceedings of the International Conference
 Foundations of the Formal Sciences V : Infinite Games, November 26th to 29th, 2004, Bonn, Germany, 
Volume 11 of  Studies in Logic, College Publications at King's
  College, London, 2007, p. 109-122. 



\bibitem[Fin01]{Fin01a} O. Finkel, Topological Properties of Omega Context Free Languages, 
 Theoretical Computer Science, Vol. 262 (1-2), July 2001, p. 669-697. 

\bibitem[Fin03]{Fin03a} O. Finkel, Borel Hierarchy and Omega Context Free 
Languages, Theoretical Computer Science, Vol. 290 (3),  2003, p. 1385-1405.

\bibitem[Fin04]{Fin04}   O. Finkel, 
An omega-Power of a Finitary Language Which is a Borel Set of Infinite Rank, Fundamenta Informaticae, 
Volume 62 (3-4),  2004, p. 333-342.

\bibitem[Fin06]{mscs06} O. Finkel, 
Borel Ranks and Wadge Degrees of Omega Context Free Languages, 
Mathematical Structures in Computer Science, Volume 16 ( 5),  2006, p. 813-840.

\bibitem[FL07]{Fin-Lec}
O.~Finkel and D.~Lecomte, 
 There Exist some $\omega$-Powers of Any {B}orel Rank.
 In the  Proceedings of the 16th EACSL Annual International Conference
  on Computer Science and Logic, CSL 2007, Lausanne, Switzerland, September
  11-15, 2007, Lecture Notes in Computer Science, Volume 4646, Springer, 2007, p. 115-129.



\bibitem[FL08]{Fin-Lec2}
O.~Finkel and D.~Lecomte, 
 Classical and Effective Descriptive Complexities of omega-Powers, 
preprint, 2008, available from http://fr.arxiv.org/abs/0708.4176.

\bibitem[HU69]{HopcroftUllman79} J.E. Hopcroft and J.D. Ullman, 
Formal Languages and their Relation
 to Automata,
 Addison-Wesley Publishing Company, Reading, Massachussetts, 1969.

\bibitem[Kec95]{kec} A.S.  Kechris, Classical Descriptive Set Theory, Springer-Verlag, 1995. 

\bibitem[KMS89]{kms}  A. S. Kechris,  D. Marker, and R. L. Sami, $\Pi_1^1$ Borel Sets, 
The Journal of Symbolic Logic, Volume 54 (3), 1989, p. 915-920. 

\bibitem[Kur66]{ku} K. Kuratowski, Topology, Vol. 1, Academic Press, New York 1966.

\bibitem[Lec01]{Lecomte01} 
D. Lecomte, Sur les Ensembles de  Phrases  Infinies  Constructibles  
a Partir d'un  Dictionnaire sur un   Alphabet  Fini, S\'eminaire d'Initiation a l'Analyse, 
Volume 1, ann\'ee 2001-2002. 

\bibitem[Lec05]{Lecomte05} 
D. Lecomte, Omega-Powers and Descriptive Set Theory, 
Journal of  Symbolic Logic, Volume 70 (4), 2005,  p. 1210-1232.

\bibitem[LT94]{lt}   
H. Lescow and W. Thomas, Logical Specifications of Infinite Computations,
 In: ``A Decade of Concurrency" (J. W. de Bakker et al., eds), Springer LNCS 803 (1994), 583-621.
 

\bibitem[Mos80]{mos} Y. N. Moschovakis, Descriptive Set Theory, North-Holland, Amsterdam 1980.


\bibitem[Niw90]{Niwinski90} D. Niwinski, Problem on $\om$-Powers posed  in the Proceedings 
of the 1990 Workshop ``Logics and Recognizable Sets"  (Univ. Kiel).

\bibitem[PP04]{pp} D. Perrin and J.-E. Pin, Infinite Words, Automata, Semigroups, Logic and Games, Volume 141 of 
Pure and Applied Mathematics, Elsevier, 2004. 



\bibitem[Sim92]{Simonnet92} P. Simonnet, Automates et Th\'eorie Descriptive, Ph. D. Thesis, 
Universit\'e 
Paris 7, March 1992.

\bibitem[Sta86]{stac} L. Staiger, Hierarchies of Recursive $\om$-Languages, Jour. Inform. 
Process. Cybernetics  EIK 22 (1986) 5/6, 219-241.


\bibitem[Sta97a]{sta} L. Staiger, $\om$-Languages, 
Chapter of the Handbook of Formal Languages, Vol 3,
 edited by G. Rozenberg and A. Salomaa, Springer-Verlag, Berlin, 1997.

\bibitem[Sta97b]{Staiger97b}   L. Staiger, On $\om$-Power Languages, 
in New Trends in Formal Languages,
Control, Coperation, and Combinatorics, Lecture Notes 
in Computer Science 1218, Springer-Verlag, Berlin 1997, 377-393.

\bibitem[Tho90]{tho}  W. Thomas, Automata on Infinite Objects, in: J. Van Leeuwen, ed.,
 Handbook of Theoretical Computer Science, Vol. B ( Elsevier, Amsterdam, 1990 ), p. 133-191.

\bibitem[Wad83]{Wadge83}
W. W. Wadge.
 Reducibility and Determinateness in the Baire Space, 
PhD thesis, University of California, Berkeley, 1983.


\end{thebibliography}
\end{document}